%% file: main.tex
\documentclass{article}

\usepackage{arxiv}

\usepackage[utf8]{inputenc} 
\usepackage[T1]{fontenc}    
\usepackage{hyperref}       
\usepackage{url}            
\usepackage{booktabs}       
\usepackage{amsfonts}       
\usepackage{nicefrac}       
\usepackage{microtype}      
\usepackage{cleveref}       
\usepackage{lipsum}         
\usepackage{graphicx}
\usepackage{natbib}
\usepackage{doi}

\title{Handling Large-scale Cardinality in building recommendation systems}

\newif\ifuniqueAffiliation
\uniqueAffiliationfalse

\ifuniqueAffiliation 
\author{ \href{https://orcid.org/0000-0000-0000-0000}{\includegraphics[scale=0.06]{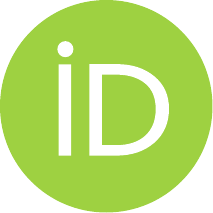}\hspace{1mm}Dhruva Dixith Kurra}\thanks{indicates equal contribution} \\
	Uber\\
	San Franciso\\
	\texttt{dkurra@uber.com} \\
	\And
         \href{https://orcid.org/0000-0000-0000-0000}{\includegraphics[scale=0.06]{orcid.pdf}\hspace{1mm}Bo Ling}\thanks{ indicates equal contribution} \\
	Uber\\
	San Franciso\\
	\texttt{bo.ling@uber.com} \\
        \And
        \href{https://orcid.org/0000-0000-0000-0000}{\includegraphics[scale=0.06]{orcid.pdf}\hspace{1mm}Chun Zh}\thanks{ indicates equal contribution} \\
	Uber\\
	San Franciso\\
	\texttt{dkurra@uber.com} \\
        \And
	\href{https://orcid.org/0000-0000-0000-0000}{\includegraphics[scale=0.06]{orcid.pdf}\hspace{1mm}Seyedshahin Ashrafzadeh} \\
	Uber\\
	San Franciso\\
	\texttt{ashahin@uber.com} \\
}
\else
\usepackage{authblk}

\newbox{\orcid}\sbox{\orcid}{\includegraphics[scale=0.06]{orcid.pdf}} 
\author[1]{%
	\hspace{1mm}Dhruva Dixith Kurra\thanks{\texttt{dkurra@uber.com}}}%

\author[1]{%
	\hspace{1mm}Bo Ling\thanks{\texttt{bo.ling@uber.com}}}%

\author[1]{%
	\hspace{1mm}Chun Zh\thanks{\texttt{chunzh@uber.com}}}%

\author[1]{%
	\hspace{1mm}Seyedshahin Ashrafzadeh\thanks{\texttt{ashahin@uber.com}}}%

\affil[1]{Uber, San Francisco}

\fi

\renewcommand{\shorttitle}{Handling Large-scale Cardinality in building recommendation systems}

\hypersetup{
pdftitle={},
pdfsubject={q-bio.NC, q-bio.QM},
pdfauthor={David S.~Hippocampus, Elias D.~Striatum},
pdfkeywords={Large scale Recommendation systems, High Cardinality, Hashing Techniques, Layer sharing, Bag of words}
}

\begin{document}
\maketitle

\input{abstract}


\input{introduction}

\input{related_work}

\input{methods}

\input{experiments_and_discussion}

\input{conclusion}

\bibliographystyle{unsrtnat}
\bibliography{references} 

\end{document}

%% file: abstract.tex

\begin{abstract}
Effective recommendation systems rely on capturing user preferences, often requiring incorporating numerous features such as universally unique identifiers (UUIDs) of entities. However, the exceptionally high cardinality of UUIDs poses a significant challenge in terms of model degradation and increased model size due to sparsity. This paper presents two innovative techniques to address the challenge of high cardinality in recommendation systems. Specifically, we propose a bag-of-words approach, combined with layer sharing, to substantially decrease the model size while improving performance. Our techniques were evaluated through offline and online experiments on Uber use cases, resulting in promising results demonstrating our approach's effectiveness in optimizing recommendation systems and enhancing their overall performance.

\end{abstract}

%% file: introduction.tex

\section{Introduction} \label{sec:intro}

\begin{figure}
	\centering
        \includegraphics[width=0.28\columnwidth, height=8cm]{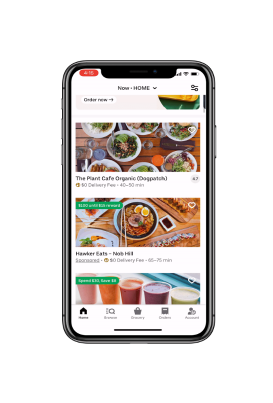}
         \includegraphics[width=0.5\columnwidth, height=8cm]{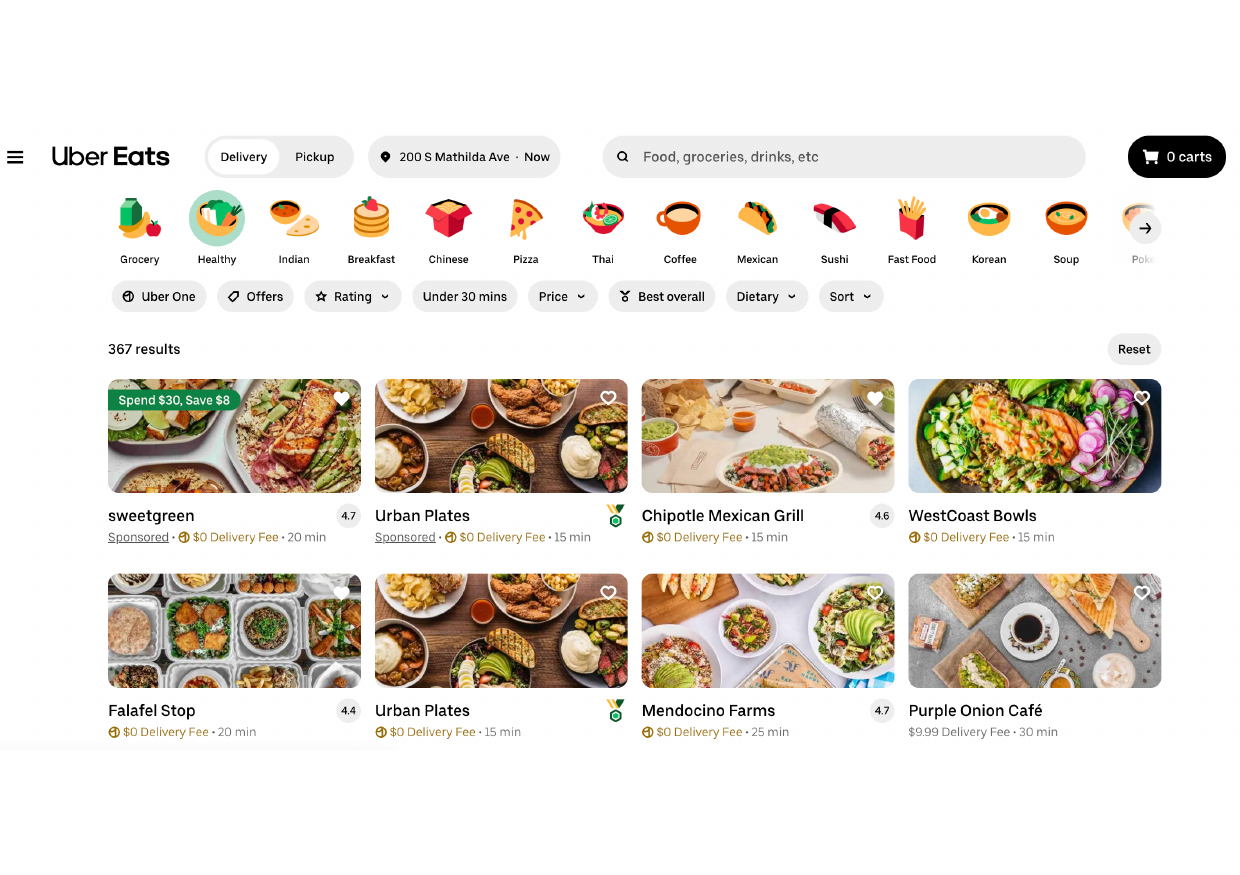}

	\caption{Uber Eats recommendation on mobile app and Website.}
	\label{fig: Uber Eats mobile app displaying personalized recommendations.}
\end{figure}
Recommendation systems have become indispensable tools for assisting users in discovering relevant items in large-scale web applications. At Uber, our recommendation systems serve a critical role in helping users discover and obtain a wide range of goods and services, from food and groceries to pet supplies and beyond (Figure \ref{fig: Uber Eats mobile app displaying personalized recommendations.}). Our recommendation systems are multi-step architectures \citep{covington2016youtube,naumov2019dlrm} that are designed to optimize specific objectives at each step. In this paper, we focus on the first pass ranker, which generates relevant candidates for subsequent rankers. This phase is particularly critical, as it is expected to achieve low latency while maintaining a high recall metric.

To improve the performance and latency of the retrieval phase, we propose new techniques that leverage high cardinal features such as $doc\_id$  or $store\_uuid$ and $query\_id$ or $eater\_uuid$. These features are typically avoided due to their impact on model size, but our research and internal experiments suggest that they can be effectively used in combination with state-of-the-art techniques to achieve improved performance. Our proposed methods have not been previously explored in the context of recommendation systems.

Our evaluation of these methods demonstrated significant improvements in performance, as measured by both offline and online evaluations, with $p\_value < 5\%$. Furthermore, we achieved a substantial reduction in model size, which is particularly important for online recommendation systems where low latency is essential.

Overall, the contributions of this paper have the potential to improve the efficiency and effectiveness of the retrieval phase in recommendation systems. By introducing novel techniques that build upon existing state-of-the-art techniques, we hope to inspire further research in this area and enable more efficient and effective recommendation systems in the future.

%% file: related_work.tex

\section{Related Work} \label{sec:related_work}

The field of recommendation systems has been extensively researched, and a variety of architectures have been proposed. To gain an understanding of the recommendation system architecture, we recommend reading \citep{naumov2019dlrm}, which provides a comprehensive overview of the topic. Typically, the recommendation system is divided into two stages: retrieval and ranking. In this research, we aim to focus on the retrieval stage of the system. Initially, retrieval systems used a static rank, such as the item's average conversion rate or PageRank \citep{5197422}, to retrieve popular items from a corpus. However, contemporary retrieval systems use a nearest-neighbor lookup in a metric space (usually $l_2$) that is customized to the query. To achieve this, the query and item must be converted into embeddings or latent factors. Until recently, matrix factorization \citep{5197422} stood as the predominant method for fulfilling this purpose. However, two-tower embeddings (TTE) have become increasingly popular since they offer the benefits of matrix factorization while also providing the ability to incorporate query and item features; further, TTE models can refine query embeddings on the fly based on context and are more flexible \citep{naumov2019dlrm}. As a result, recent research has focused on training TTE. TTE models, therefore, present a compelling solution for improving the efficiency and performance of recommender systems in large-scale applications. State-of-the-art TTE training libraries such as TensorFlow Recommenders use in-batch negatives, where the sampling bias correction is done to offset the fact that popular items are penalized more since they appear more often as negatives \citep{yi2019logq}. Additionally, researchers have explored using item embeddings as cross-batch negatives \citep{wang2021crossbatch, lindgren2021efficient} in settings where the model's actual batch size is limited by memory constraints. The concept of mixed negatives, in which a pool of negatives is randomly sampled, and another pool is taken from in-batch negatives, has also been investigated \citep{ji2020mixednegatives}. To handle the high cardinality of the features, a usual technique is to map it to a relatively smaller embedding table; hashing techniques like Compositional Embedding \citep{shi2020compositional}, and MultiHash Embedding \citep{tito2017hash} are being introduced for this purpose.

All of these techniques significantly improve the accuracy of TTE training. Our research builds on existing techniques to propose a generalized use case that could be used across Uber and beyond.

%% file: methods.tex

\section{Methods} \label{sec:methods}

\begin{figure}
    \centering
    \includegraphics[width=0.675\columnwidth]{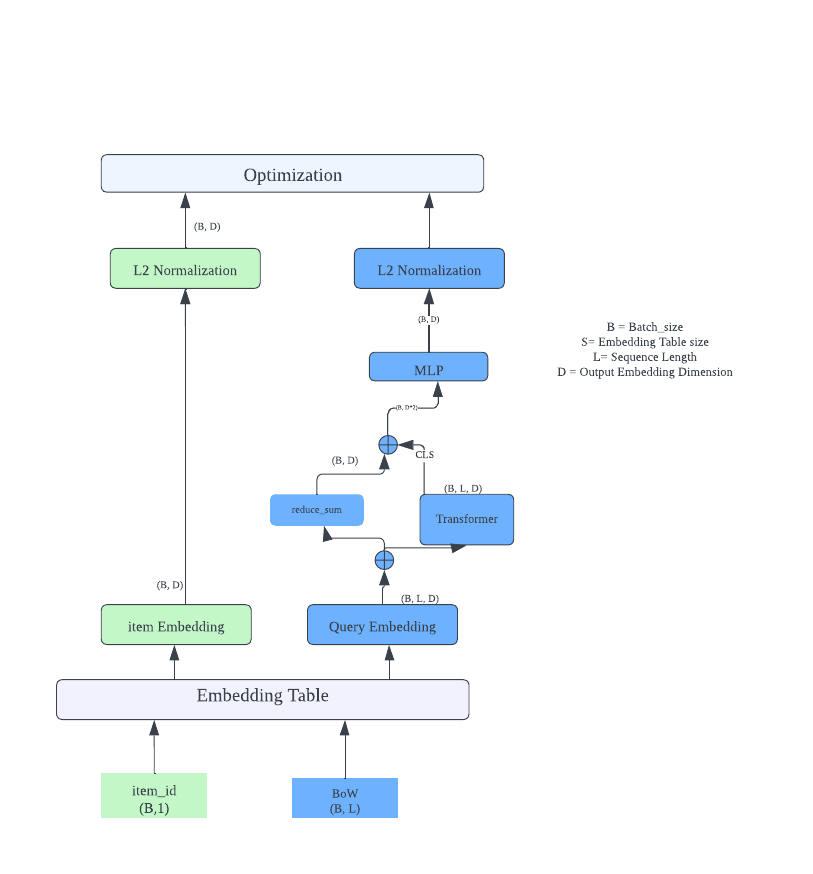}
    \caption{Simplified BoW with layer sharing architecture}
    \label{fig:Simplified BoW with layer sharing rarchitecture}
\end{figure}

The two-tower embedding models commonly used in recommendation systems are composed of query and item towers, where any input feature can be utilized. However, our ablation studies have demonstrated that UUID features, such as $eater\_uuid$ and $store\_uuid$,  play a crucial role in the system's performance, and their exclusion can negatively impact results. Unfortunately, the high-cardinality of these features poses a challenge for inclusion in the model, as it would significantly increase the model size. This, in turn, would have a negative impact on inference times, training time, and performance metrics, as the model would have difficulty learning from extremely sparse features. Existing approaches, such as hashing or QR embedding \citep{tito2017hash, shi2020compositional}, have not proven effective in our use cases.

To address this issue, we developed a bag-of-words approach, as detailed in section ~\ref{sec:bag_of_words}, which helped reduce the model size. We also utilized layer sharing, detailed in section ~\ref{sec:layer_sharing}, to further decrease model size and complexity. These techniques, in conjunction with other state-of-the-art approaches, have significantly improved the performance of our recommendation systems.

\subsection{Bag of Words (BoW)} \label{sec:bag_of_words}

\begin{figure}
    \centering
    \includegraphics[width=0.675\columnwidth]{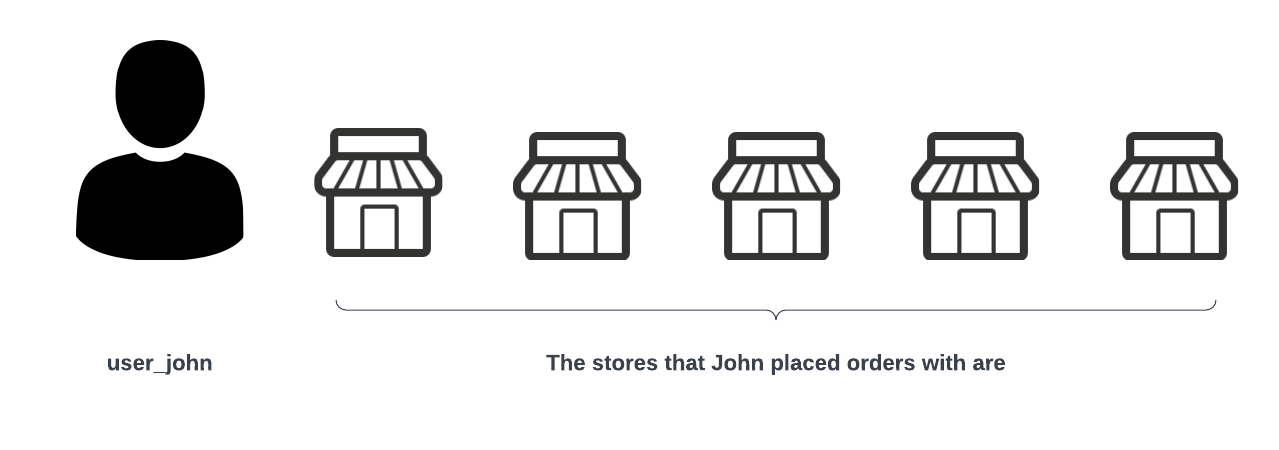}
    \caption{ Bag of Words: Using a user\'s order history with specific stores as a proxy for the user, effectively reducing the dimensionality from user\_id (several hundreds of million scale) to store\_id (million scale)}
    \label{fig:bag_of_words}
\end{figure}

For our Uber recommendation system use case, the $eater\_uuid$ and $store\_uuid$ features have incredibly high cardinality, with the $eater\_uuid$ part having a much higher cardinality than the $store\_uuid$ feature. As the number of $store\_uuid$ $<<<$ number of $eater\_uuid$, the $eater\_uuids$ are the primary contributor to the model size, which poses a significant challenge. However, if we can replace the $eater\_uuid$ with a proxy, we can solve the issue of model size.

To solve this issue, we propose replacing $eater\_uuid$ with a Bag of the user's previously ordered stores, as inspired by the YouTube paper\citep{covington2016youtube}. This Bag serves as a proxy for the users’ UUID and captures the user's behavior based on previously ordered stores. Bag of Words approach is illustrated in Figure \ref{fig:bag_of_words}

The BoW feature offers several advantages. First, the BoW model's size is much smaller than the previous approach where we were able to reduce the model size by 25x. Second, training time is reduced from days to a few hours, resulting in significant time savings. Finally, the BoW model requires less computation, making it more efficient and easier to maintain.

\subsection{Layer Sharing} \label{sec:layer_sharing}

Traditional two-tower models for recommendation systems keep the query and item towers separate, with no sharing of input features between them. TTE models provide a significant advantage as the input features from the query tower are not used in the item tower and vice versa. This characteristic allows for joint training of the model while still enabling the separate deployment of each tower, reducing latency in serving. In contrast, our proposed TTE model introduces a novel technique inspired by transformer architecture and AlexNet \citep{krizhevsky2017imagenet}. We enforce embedding table-layer sharing for UUID features between the two towers, allowing them to share information and capture more nuanced relationships between the query and item. While this technique is not typical in most textbook TTE models, we have found it to be essential in significantly improving our model's performance.

By sharing the embedding table layer, the TTE model reduces the number of parameters, which in turn reduces the model's size and inference time, making it more computationally efficient. Additionally, by allowing the towers to exchange information about the high cardinal features, we can capture more subtle relationships between the query and item towers, leading to more accurate recommendations.
Overall, these two additions to our TTE model have proven to be powerful techniques in our recommendation system. It demonstrates the potential for innovative adaptations of established models to achieve significant improvements in performance.

\subsection{Baseline Models} \label{sec:baseline_models}

This study compares three different recommendation models: the Deep Matrix Factorization (baseline) model, and two variations of the bag-of-words (BoW) model. All models incorporated optimization techniques, including logQ correction, removal of accidental hits, cross-batch negatives, and softmax temperature correction. While the TTE model allows for the addition of features, we did not incorporate any additional features to maintain a fair comparison with the baseline model.

\begin{itemize}

\item \textbf{Deep Matrix factorization (baseline) model}
A city\-level model, inspired by the YouTube paper \citep{covington2016youtube} and CBN \citealp{wang2021crossbatch}, utilized the $nn.Embedding$ module to encode user and store UUIDs. We optimized the DMF model using AdaGrad with a learning rate of beta 1.0 and a $cache\_size$ of 6144, with a batch size of 8192, cross-entropy as a loss function, and embeddings of size 32. The model consisted of an additive layer for query and item towers, respectively, with a dot product output, trained for 10 epochs.

\item \textbf{Bag of Words model (BoW Model)}
The BoW model is a single global model with the same input features as the baseline model, replacing the $user\_uuid$ with a bag-of-words approach as explained in section ~\ref{sec:bag_of_words}. We optimized this model using AdaGrad with a learning rate of 0.02, beta 1.0, and a cache size of 6144, with a batch size of 8192 and embeddings of size 32.

\item \textbf{Bag of Words model with layer sharing}
This variation of the BoW model incorporated layer sharing, as explained in section ~\ref{sec:layer_sharing}, with the same hyperparameters as the BoW model. The architecture of BoW with Layer Sharing is illustrated in Figure \ref{fig:Simplified BoW with layer sharing rarchitecture}
\end{itemize}

\subsection{Evaluation Metrics} \label{sec:evaluation_metrics}

The evaluation of recommender system models is a crucial aspect of developing effective recommendation algorithms. In this study, we used multiple evaluation metrics, including Hit Rate (Recall), Normalized Discounted Cumulative Gain (NDCG), and Mean Average Precision (MAP), to evaluate our models.

We generate embeddings by making predictions of the trained two-tower model and intercepting the output of the user and item towers. Then we used these user and item embeddings to compute factorized-top-$k$ items for every user and compute Hit Rate (Recall) based on previous user-item interactions. 
\begin{equation}
    Hit\ Rate@k = (number\ of\ relevant\ items\ retrieved\ in\ top\ k)\ /\ (total\ number\ of\ relevant\ items)
\end{equation}

We used Hit Rate@{5, 20, 100, 200, 300, 400, 500} as one of our evaluation metrics and Hit Rate@500 as our optimizing metric, and the other values of k as satisfying metrics. We chose Hit Rate as our primary metric since in the first stage of the multi-pass architecture, we care mostly about the returned list containing as many relevant items as possible even though the ordering or scores of those items may not be aligned with other stages.

By using multiple evaluation metrics, we were able to gain a comprehensive understanding of the performance of our models and their ability to provide accurate recommendations to users.

%% file: experiments_and_discussion.tex
\section{Experiments and Discussion} \label{sec:experiments_and_discussion}

In this section, we present the results of our offline experiments to evaluate the performance of the three proposed models. The training data utilized for the experiments consisted of four months of Uber Eats order data, which was randomly sampled on 30\% of users. The order data can originate from any source, such as a home feed, carousel, or search. We used one week of data for validation and the remaining data for training.

To provide each user with a list of their previously ordered stores, we employed a bag-of-words approach mentioned in section ~\ref{sec:bag_of_words}. Specifically, we collected the last 120 days of previously ordered stores into a time-sorted list of $store\_uuids$. The models were trained using PyTorch\textsuperscript{\tiny\textregistered} and TorchScript on Michelangelo, Uber's internal machine learning platform. PyTorch\textsuperscript{\tiny\textregistered} was chosen as the framework due to its extensive support for deep learning at Uber.

\begin{figure}[h]
    \centering
    \includegraphics[width=0.675\columnwidth]{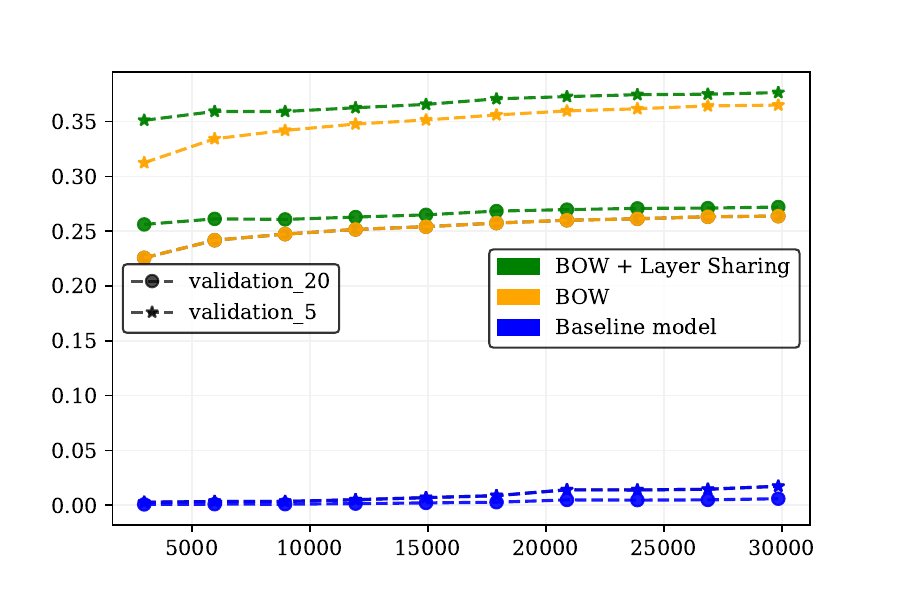}
    \caption{Comparison of Validation Metrics between Baseline Models and BoW Models at lower hit rates\{5, 20\}. The BoW models outperform the baseline models by a significant margin, as indicated by the hit rate@\{5, 20\} values
}
\label{fig: hit_rates}
\end{figure}

For evaluating the models, we used one week of order data as a test dataset and evaluated the models based on the metrics described in section ~\ref{sec:evaluation_metrics}. The results of our experiments are presented in Figure \ref{fig: hit_rates} and Table \ref{tab:metirc_recall_higher} 

Our analysis of the results highlights the superior performance demonstrated by the newly proposed Bag of Words (BoW) models across all recall values in comparison with the Baseline model. We would like to highlight that the Baseline model encountered challenges related to convergence and prolonged training hours, primarily stemming from the considerable model size induced by large Embedding Tables.

Our examination of the results highlights the superior performance of the newly proposed Bag of Words (BoW) models across all recall values compared to the model Base Model. Notably, our key metric, Recall@500, experienced a substantial improvement of 10\%. Furthermore, the lower recall values (k=100, 200) showcased a noteworthy enhancement, with a 30\% increase in recall.

\begin{table}
    \centering
	\resizebox{0.95\columnwidth}{!}{
	\begin{tabular}{lrrrrr}
		\hline
		Model & Hit Rate@100 &  Hit Rate@200 &  Hit Rate@300 &  Hit Rate@400 &  Hit Rate@500 \\ \hline\hline
		without BoW & $0.4858$ & $0.634$ & $0.7198$ & $0.7766$ & $0.8165$ \\ \hline
		BoW & $0.7328$ & $0.8275$ &  $0.8752$ &  $0.9041$ &  $0.9232$  \\ \hline
		BoW with Layer Sharing & $\textbf{0.7334}$ &  $\textbf{0.8286}$ &  $\textbf{0.8763}$ &  $\textbf{0.9051}$ &  $\textbf{0.9243}$ \\ \hline
	\end{tabular}}
	\caption{Performance evaluation for the baseline model and bag of words model, bag of words model with layer sharing. Recall@K is calculated for different values of K to simulate our retrieval systems. For each metric, the highest performance is shown in boldface.}
	\label{tab:metirc_recall_higher}
\end{table}

\subsection{Convergence speed}

\begin{figure}
    \centering
    \includegraphics[width=0.675\columnwidth]{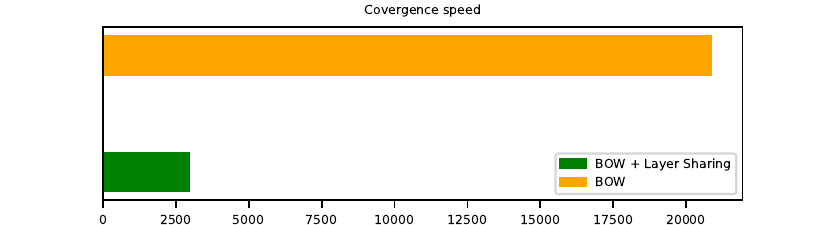}
    \caption{BoW model with layer sharing converges faster (2000 steps) to hit rate@20 threshold of 0.35 compared to simple BoW model (20k steps), indicating improved training speed and performance.
}
    \label{fig:convergence_rate}
\end{figure}

In this study, we evaluated the convergence speed of the Bag-of-Words (BoW) model with layer sharing and the simple BoW model. We found that the BoW model with layer sharing achieved convergence to the hit rate@20 threshold value of 0.35 within 2000 steps, whereas the simple BoW model took around 20k steps to reach the threshold (Figure \ref{fig:convergence_rate}). This indicates a significant improvement in training speed for the BoW model with layer sharing, while also being more performant than the simple BoW model

The improved performance of our proposed BoW models can be attributed to the ability of the models to leverage the temporal ordering of the previously ordered stores, which allows for a more accurate representation of the user's ordering behavior. The layer sharing of the BoW models also contributes to superior performance by allowing for the sharing of information between the two towers, which leads to more efficient learning and improved convergence speed.

Overall, the results of our experiments demonstrate the effectiveness of our proposed BoW models and their ability to improve the recommendation accuracy of the Uber Eats platform.

%% file: conclusion.tex

\section{Conclusion} \label{sec:conclusion}

We focused on addressing the challenge of handling high-cardinality features, such as UUID, in building recommendation systems. The proposed Bag of words approach and layer-sharing technique were introduced as innovative solutions to this problem, resulting in a significant reduction in model size, improved performance, and convergence speed, as demonstrated by offline and online evaluations. Our work contributes to the development of more efficient and effective recommendation systems, particularly in the retrieval phase. By building on existing state-of-the-art techniques, the proposed methods have the potential to inspire further research in this area, with the ultimate goal of enabling more efficient and effective recommendation systems in the future.